# Hydromechanical considerations on the origin of the pentaradial body structure of echinoderms

Michael Gudo[1]

## Abstract

When echinoderms are conceptualized as hydraulic entities, the early evolution of this group can be presented in a scenario which describes how a bilateral ancestor (an enteropneust-like organism) gradually evolved into a pentaradial echinoderm. According to this scenario, the arms are outgrowths from the anterior/posterior body axis of the bilateral pterobranchia-like intermediate. These outgrowths developed when the originally U-shaped mesentery of the intestinal tract formed loops, and correspondingly, the tensile chords of the mesentery were attached to the body wall in five loops. The wall faces between these regions of tensile chords could bulge out under the hydraulic pressure of the body coelom. The originally more or less round body cavity was deformed into a pneu with five bulges. The loops of the gut forced a roughly symmetric arrangement, which was enhanced by a physical fact: five pneus as well as one pneu with five internal tethers, naturally adopt a pentaradial pattern of 'minimum contact surfaces', as the most economic arrangement. This part of the body became shorter along its longitudinal axis, but gained volume by the five bulges. The bulges came in contact with the tentacular crown, placed directly in front of them, and began to grow under the tentacle bases. The more the bulges extended along the tentacles, forming the so-called arms, the better was the mechanical support for the tentacles, which are completely integrated in the arms. Only the paired end-branches of the tentacles project and the pentaradial ambulacral system with many ambulacral podia was the result. These evolutionary transformations were accompanied by certain histological modifications, such as the development of mutable connective tissues and skeletal elements that fused to ossicles and provided shape stabilization in the form of a calcareous skeleton in the tissues of the body wall. The resultant organism was an ancestral eleutherozoan echinoderm (Ur-Echinoderm) with an enlarged metacoel stabilized by hydraulic pressure working against a capsule of mutable connective tissue, skeletal elements and longitudinal muscles.

## Introduction

Recent eleutherozoan echinoderms show a pentaradial symmetry of their body shape and of their internal organ systems. This symmetric arrangement develops ontogenetically by a specific growth process that is well known from embryological investigations (David & Mooi 1998; Hart 2002; McCain & McClay 1994; Morris 1999; Wray 1997). However, the evolutionary origin of the pentaradial organisation is almost unknown. Several ideas have been presented to explain the origin of pentaradial symmetry in echinoderms (David & Mooi 1999; Hotchkiss 1997; Jefferies 1991; Kerr & Kim 1999; McCain & McClay 1994). Most of these authors argue about the efficiency and advantages of the pentaradial organisation of the tentacles of sea lilies, but none of them offers a mechanism by which the origin of the entire pentaradial organization (comprising the outer body shape and internal organ systems) is explained sufficiently. Peterson (2000a) argues that the anterior/posterior axis (A/P-axis) was preserved throughout the entire evolutionary pathway of the echinoderms and that the arms and therefore the pentaradially symmetric bauplan of echinoderms developed by outgrowths of this original central body axis. In support he draws three lines of evidence: the expression patterns of a posterior class Hox gene in the coeloms of the nascent adult larvae, the anatomy of some early fossil echinoderms, and the relation between endoskelatal plate morphology and the associated coelomic tissues. All indicate that the anterior/posterior body axis runs from the mouth through the

1) Dr. Michael Gudo, Morphisto – Evolutionforschung und Anwendung GmbH (Institut für Evolutionswissenschaften), Senckenberganlage 25, 60325 Frankfurt am Main, Germany, e-mail: michael.gudo@morphisto.de

adult coelomic compartments, and consequently there is but a single plane of symmetry dividing the echinoderms into left and right halves. Peterson's arguments are quite useful, however, two simple, but crucial questions remained to be answered: (1) how was the pentaradial symmetric arrangement attained? and (2) what did the ancestral and intermediate stages look like: i.e. how were organisms that represent these stages structured and organized?

## Theoretical Background

To answer both of these questions, a hydraulic conceptualization of echinoderms can provide insights into the internal organization and the evolutionary process that must have taken place. Following the structural-functional approach of engineering morphology (Gudo et al. 2002) echinoderms can be analyzed in a way similar to that used by an engineer to analyse any technical apparatus which he does not know. Therefore the arrangement and biomechanical coherence of hydraulic cavities, anatomical structures (= the structural-functional organisation, functional design or „Körper-Konstruktion")ized, the processes of form generation and form preservation during individual development and evolutionary transformations are in the focus of interest (= engineering morphology, Gudo 2002; Gudo, et al. 2002). If echinoderms are investigated as dynamic, energy transducing machines and hydraulic entities one has to conclude that they are functional units which could not evolve arbitrarily. Constrained by structural-functional, biomechanical and hydraulic principles their evolution has to follow specific paths (anagenetic pathways) which determine a kind of morphospace of those evolutionary changes which result in viable body structures (Gutmann 1988, 1995). Therefore research on echinoderm evolution has to deal with two major aspects: (1) Conceptualization of an ancestral echinoderm as a structural-functional entity and (2) the presentation of a unidirectional evolutionary scenario, i.e. a historical evolutionary theory (sensu Bock 1991; Bock 2000; Gudo & Grasshoff 2002) that reconstructs the anagenetic transformations of the functional designs of their soft-bodies.

## Conceptualization of echinoderms as hydraulic entities

In the echinoderm literature, it is generally assumed that the ambulacral system acts as a hydraulic system. As can be observed in any living sea-star or see urchin, each podium is protruded when hydraulic pressure in the ampullae is increased by contraction of ampullar muscles and it is retracted by longitudinal muscles in the wall of the tube (compare Woodley 1967). Thus the fluid filling of the ambulacral system acts as a hydraulic skeleton and transmits the forces of the muscles. In precisely the same manner the coelomic fluid of the body cavity works as a hydraulic skeleton for the muscles of the body wall. When a sea-star, for example, moves its arms, the fluid filling in this arm acts as a force transmitter. Also during peristaltic movements of a sea-cucumber the coelom filling works as a hydraulic skeleton for the muscles and mutable tissues in the body wall. As these examples show, it is appropriate to describe the entire body of echinoderms as a pneu (Gutmann 1981, 1988). The influences of hydraulic shape-determining mechanisms during growth have already been described for sea-urchins (Dafni 1984, 1986, 1988). Accordingly, the body shape is the result of the interaction of the muscles, mutable connective tissues and skeletal elements via the hydraulic fluid of the coelom filling. Although some authors have neglected the hydraulic shape-determining mechanisms that are valid for sea urchins (Nachtigall & Philippi 1996), it stands beyond any necessary discussion that every fluid filling which is enclosed by muscles, connective tissues and skeletal elements is permanently or at least temporarily under pressure and therefore acts as a hydraulic fluid. Therefore the coelomic cavity not only determines the body shape, but with all surrounding anatomical structures it is a functional complex that limits the scope of evolutionary changes, because pneus can only be transformed as functional units (there are no half pneus) (Gudo 2004).

### Choice of an ancestor

Before the anagenetic reconstruction can start, it is necessary to select an ancestor as a plausible starting point for the evolutionary pathway. This choice of the ancestor is determined on the results of other investigations, such as traditional morphological or embryological approaches, molecular results or palaeontological findings. These indicate that enteropneusts and pterobranchs are sistergroups of echinoderms (Bromham & Degnan 1999; Halanych 1995; Janies 2001; Lowe & Wray 1997; Smith et al. 1993; Turbeville et al. 1994; Wada & Satoh 1994) and therefore these organisms are eligible to use them as a starting point for reconstructing an evolutionary pathway. Nevertheless, other organisms have been proposed as ancestors for the echinoderms (Garstang 1928; Gislén 1930; Grobben 1923; Jollie 1962; Nichols 1962, 1967) and echinoderms have even been considered as ancestors to the chordates (Eaton 1970; Jefferies et al. 1996). Summarizing all models which have been presented before, every deuterostome and even some protostome animals have already been mentioned as ancestor for echinoderms (for a summary compare Gudo & Dettmann 2005, in press). However, structural-functional, morphological and molecular data are not consistant with most of these proposals, so that an enteropneust-like organism can be seen as the most plausible ancestor for the echinoderms (Cameron et al. 2000; Gutmann 1969, 1973; Peterson et al. 2000b).

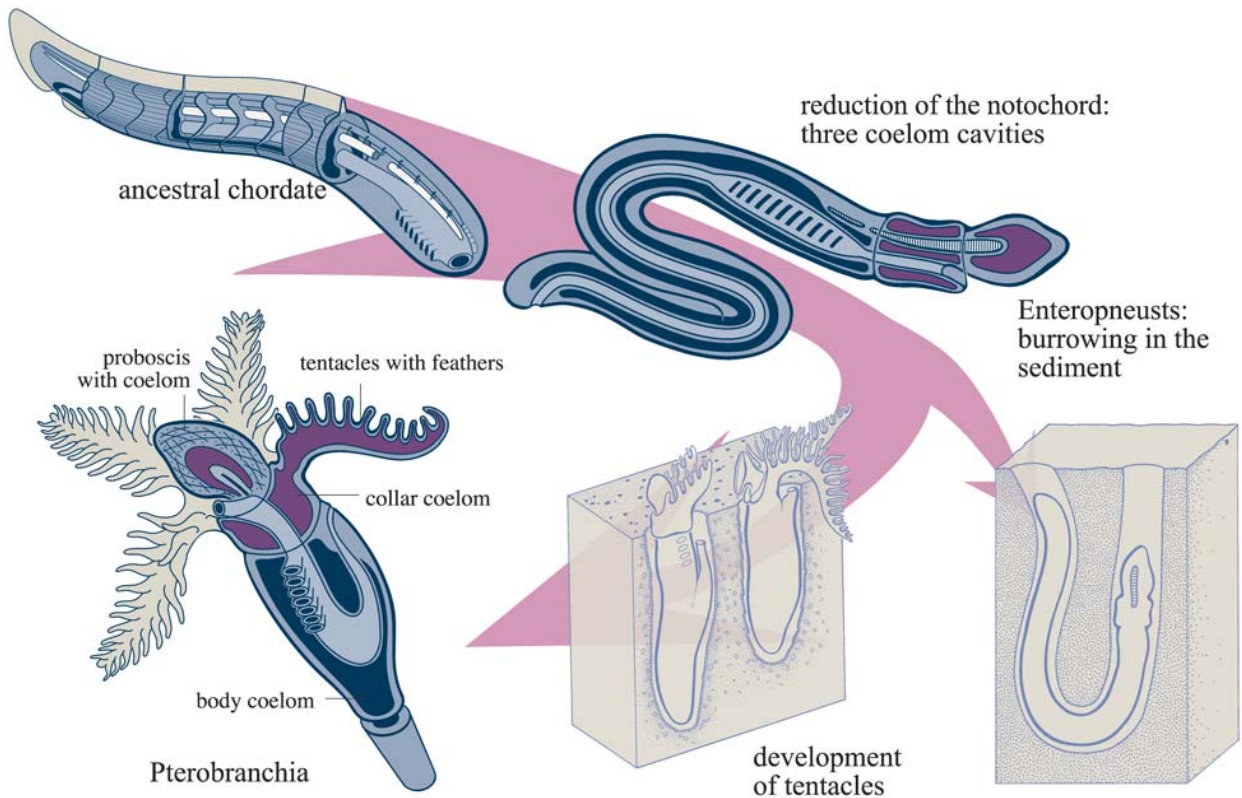

**Text-fig. 1.** Early chordates are assumed to have had longitudinal muscles and a notochord to preserve the body length and to have moved by lateral bending of their body. As seen in recent Branchiostoma, chordates with such an organization can also burrow in the sediment. If the notochord reaches over the mouth region, evolutionary modifications are possible, by which burrowing was improved. Accordingly an enteropneust-like organism evolved in which the notochord was reduced and at the front end a proboscis developed which was capable of peristaltic movements. The region behind the proboscis formed a collar which was necessary for burrowing by peristaltic movements of the proboscis. The hind end of the body remained elongated but the body wall contains only longitudinal muscles. From this organization on the one hand the enteropneusts branched and on the other hand the formation of tentacles and bending of the gut lead to the pathway of pterobranchs. From this evolutionary the echinoderms evolved by further differentiations (see Figure 3).

### Reconstruction of an evolutionary scenario

An evolutionary scenario is a historical narrative which in the case of echinoderms attempts to explain how pentaradial organization evolved from the bilateral organization of the pterobranch-like ancestor and how the enormous enlargement of the body cavity was achieved. For consistency of model presented here, the evolutionary history of the ancestor is also considered. This also provides a more complete representation of the anagenetic transformations which can finally support or challenge supposed phylogenetic relationships (Huxley 1957; Peters 2002).

### Evolution of the ancestor

The bilaterally symmetrical ancestor of the echinoderms is envisaged to have evolved from a chordate-like deuterostome ancestor in which the notochord extended over the mouth opening. Such organisms were able to adopt a burrowing life-style in the sediment and, during further evolution, the muscle fibres of the front end crossed over each other and formed a three-dimensional muscle grid capable of peristaltic movement thereby improving their ability to burrow. As locomotion was no longer driven by lateral bending of the body, the notochord and the longitudinal muscles were no longer necessary and became reduced as far as possible. At the front end, the proboscis and a muscular collar surrounding the mouth opening remained as thicker muscle grids. In this region also the stomochord remained as a stabilizing structure. In the hind end (the metasome) only a thin layer of longitudinal muscles remained. These muscles have no direct antagonism. They are passively stretched when the animal moves through the sediment. During such movements, the metasome is pulled afterwards occasio-nally (for more details compare Gudo & Grasshoff 2002; Gutmann 1981, 1988; Peters & Gutmann 1972).

From this enteropneust-like ancestor the step to a pterobranch-like organism is quite small. Simply by developing tentacles from the collar a quasi sessile life-style can be attained. Tentacles can be used to catch nutrition from the water column, and therefore these organisms were able to exploit new environmental conditions. This can be seen from a structural-functional point of view as a process of economization when in further evolutionary transformations the hind opening of the gut was shifted to the front end, outside of the tentacle crown, a situation already attained in pterobranchs; the mesentery with the intestinal tract attained an U-shaped course (Figure 1). The metasome now consisted of two parts, the anterior part with the U-shaped gut and a fluid filled posterior part in which only thin longitudinal muscles persist in the body wall. Basically the same muscle interaction as mentioned for enteropneusts can be observed in living pterobranchs. When a pterobranch creeps along a substrate, the longitudinal muscles of the trunk are stretched; when these muscles contract, they pull the body back. As in enteropneusts, the longitudinal muscles of pterobranchs do not have a direct antagonism. However, an agonistic-antagonstic relation exists between the proboscis and the longitudinal muscles of the trunk. This relation is of particular importance for further evolutionary possibilities, because it constrains the sessile life style and the evolution of larger feeding apparatuses.

There were two subsequent possibilities for bending the intestinal tract. On the one hand the U-shaped course can be attained by shifting the anus in the mesenterium from caudal to rostral, so that the gut and the mesenterium come to lie in one plane. This will open the evolutionary possibilities seen in the pterobranchs, because this configuration of the intestinal tract allows the body to become smaller; hereby it is even possible that the mesentery was entirely reduced. On the other hand the U-shaped course can be attained by bending the gut together with the mesentery until finally two parallel planes are formed. According to this organisation, the internal tensile chords confine body enlargement to the orginal dorsal-ventral plane. The result is an inflated body shape which is slightly compressed laterally, and it seems likely that from such an evolutionary stage the 'homalozoans' sensu lato branched off.

From these initial morphological transformations it can be concluded that all echinoderms owe their existence to particular transformations which are characterized by hydraulically inflated bodies. Consequently the body shape could only be preserved if additional shape stabilizing anatomical structures developed, because the remaining longitudinal muscles were not sufficient. The structures which must have developed at this stage are mutable connective tissues, skeletal elements and further internal or dermal tensile chords. These structures are capable of generating a constant pressure on the coelom fluid which therefore acts a hydraulic skeleton. In terms of evolutionary explanations the formation of mutable connective tissues with skeletal elements represents an economization, because these tissues need less energy than muscles and their contractility is much more efficient.

The inflation of the body led to the formation of two distinct parts of the body, a voluminous anterior part and a thin posterior part, the trunk. As a consequence the intestinal tract has to be looped, if it were not to have been dramatically shortened. Since the intestinal tract is mechanically attached to the body wall via the connective fibres of the mesentery, the possibilities of transformations are limited. One possibility is to develop one addtional loop, and the second possibility is to develop two additional loops. Accordingly the collagenous fibres of the mesentery follow these loops and reach the body wall at several places. Where the fibres are connected to the body wall they work as internal tensile chords and restrain further inflation.

Attaining this stage was of crucial importance for the entire subsequent evolution of echinoderms, because numerous evolutionary opportunities were opened up. However, here is not the place to derive all the echinoderm pathways. The fossil record shows that pentaradiate and triradiate echinoderms evolved quite early. This speaks for a minimum of three evolutionary pathways of echinoderms, two for pentaradial echinoderms (one with direct development, one with indirect development) and one for asymmetric echinoderms. I will focus here only on the pentaradial echinoderms.

## Pentaradial echinoderms

Pentaradial echinoderms arose when the body cavity was widely inflated and the mesentery with the intestinal tract formed two additional loops (U-shaped bows). Initially the U-shaped intestinal tract was rotated about 90 degrees and then from the lower part two new loops developed laterally. Accordingly the tensile chords of the mesentery shifted along the inner surface of the body coelom. Finally connective fibres of the mesentery provided tensile chords in five regions of the anterior body part, and these fibres limited further inflation to those places on the wall faces where fibres were absent. The latter, not held by the mesenterial tensile chords, bulged out under the pressure of hydraulic fluid and when these bulges attained a sufficiently large size, a physical principle (the principle of the minimum contact surface of hydraulic pneus) forced them into a permanent pentaradial symmetry (Figure 2).

This evolutionary transformation represents a nomological-deductive explanation (i.e. a functional explanation, sensu Bock 1991; Bock 2000), whereby the pentaradial arrangement of regions with and without tensile chords is a consequence of the certain mode of elongation of the intestinal tract bringing tethering structures in five regions to the body wall. However as already mentioned before – there are two pathways of pentaradial echinoderms: the five pneus develop either directly or indirectly; Figure 2 shows the indirect way; however in the direct way steps (B) and (C) are omitted.

All the intermediate stages are functional and their advantage in selection can be seen in the elongation of the intestinal tract which provided for better digestion. The transversal enlargement of the body coelom (metacoel) lead to another crucial transformation: the tentacles were first mechanically supported from below and finally they were enclosed by the body tissues until only their pinnulate ends projected over the hydraulic bulges. The tentacles were transformed into the ambulacral system and the hydraulic bulges of the body are now the so-called arms. These arms were stabilized by the

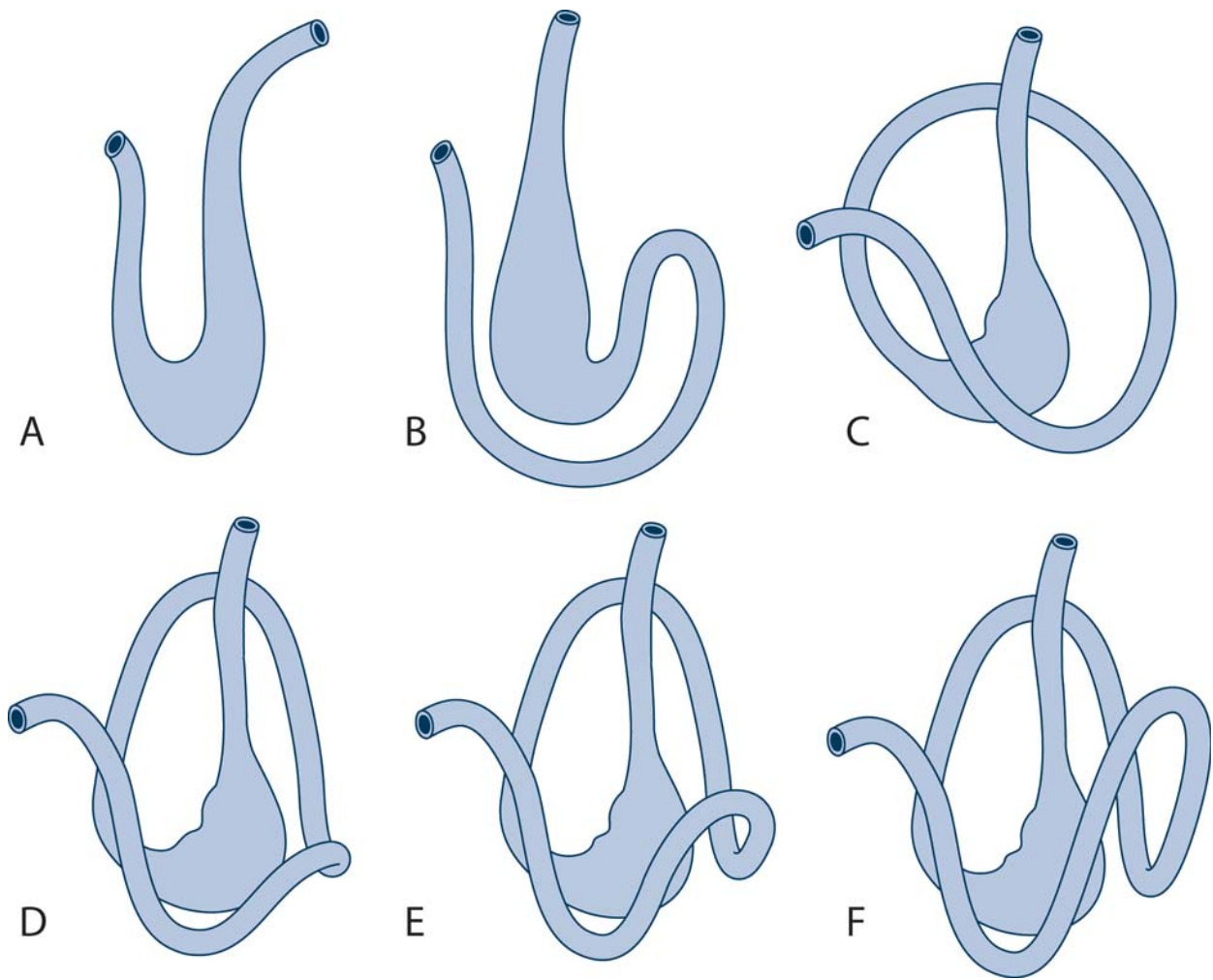

**Text-fig. 2.** It is proposed that the echinoderm ancestor evolved directly from a bilateral pterobranch-like ancestor with an U-shaped gut, when the trunk was enlarged in diamter and shortened in its length. Hereby the intestinal tract was lengthened. The original U-shaped intestinal tract (A) formed loops. The sequence from (B) – (F) shows how concomitantly to the elongation of the intesinal tract firth one and then two additional loops were formed. Since the intestinal tract in connected to body wall via the fibrous mesentery, connective tissues fibres follow this new course of the intestinal tract and work as internal tensile chords (tethering structures) which limit the further inflation of the body wall. In a physical sense this situation can be understood as one pneu with five internal tetherings and such a attains the same symmetric arrangements as five individual pneus which for physical reasons arrange themselves with minimum contact surfaces in a pentaradial pattern.

internal pressure, working against the stiffening of mutable connective tissues and skeletal elements. Longitudinal muscles from the body wall reached into the arms. The fluid filling of the arms works as a hydraulic skeleton so that the arms could be moved (Figure 3).

The ambulacral system developed in several steps. The first involved the formation of a groove in each arm in which the tentacle lay. In the second step dermal tissues grew around the tentacles, until only the pinnulated branches (=podia) projected of the body tissues. Small ampullae (one on each podium) developed and reached into the coelom of the arms to provide a fluid reservoir for the podia to be protruded and retracted. The tentacles, which developed originally from the collar, still had the fluid filled collar coelom (mesocoel) which worked as an independant hydraulic system.

The ambulacral system consists of the central ring-canal and its five branches, the radial canals and the pinnulate branches which form the numerous podia. From the ring canal the so-called stone canal connects the ambulacral coelom with the outer medium. This stone canal is a remnant of the original proboscis with its coelomic cavity (protocoel) and hydropore.

The body structure which was attained by these transformations is the pentaradial Ur-Echinoderm. It has pentaradial organisation of its internal anatomy, an ambulacral-system which is enclosed in the body coelom, and five hydraulic bulges (arms) which were capable of capturing nutrition from the water column. From this Ur-Echinoderm all the other pentaradial echinoderms evolved by further differentiations.

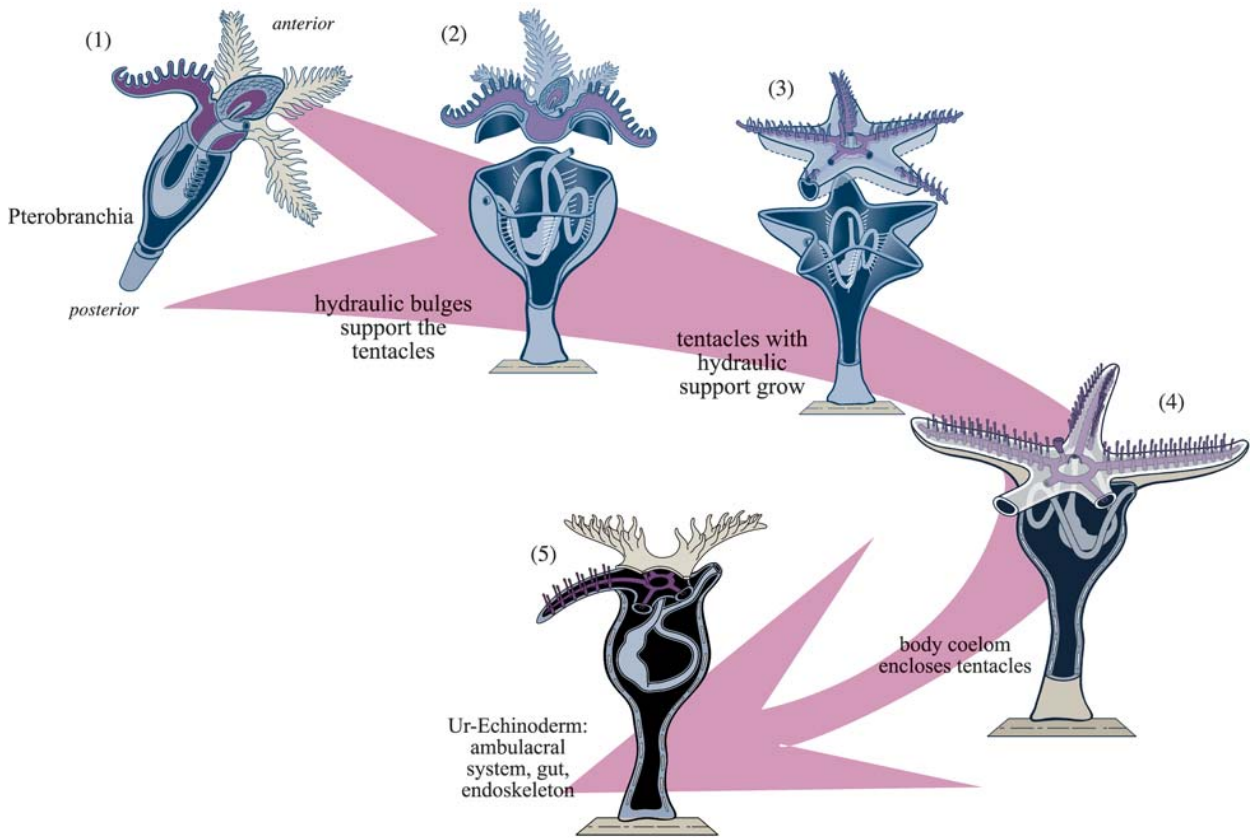

**Text-fig. 3.** From the trunk (metasom) of the bilateral ancestor (1) hydraulic bulges grow out in those regions where no tensile chords of the intestinal tract made contact with the body wall (2). These are the regions of hydraulic pneus, as shown in Figure 2. These outgrowths of the anterior/posterior axis support the tentacles which had already developed from the collar, so that these tentacles could become larger. Those tentacles which did not get such hydraulic support were reduced. Since only five hydraulic outgrowths could develop (see Figure 2) only five tentacles could be supported hydralically the pentaradial pattern of the intestinal tract was moulded onto the entire body. Since tentacles and hydraulic bulges were in close contact to each other (3), the radial canals of the tentacular crown were countersunk into the dermal tissues, but from each podium one end looked out of the tissues to the outer medium and one end (the ampullae) looked out of the tissues into the trunk coelom (4). The ambulacral-system has been developed. (5) Finally a body structure results that is typical for an echinoderm; this Ur-Echinoderm opened up the wide evolutionary field of the pentamerous echinoderms, in which the course of the intestinal tract could undergo various modifications.

## Discussion

The evolutionary transformation of a bilateral enteropneust-like ancestor into the Ur-Echinoderm was accompanied by certain histological and gross anatomical transformations (Lowe & Wray 1997). In this context, the original A/P body axis was preserved, while the arms developed from outgrowths of the central body. Even in Recent echinoderms, the oral-aboral axis corresponds to the A/P-axis of the bilateral ancestor, as suggested by molecular and morphological evidence (Peterson, et al. 2000a). However, the crucial morphological transformation have been detailed here.

Nevertheless, further aspects are of interest within the evolutionary history of echinoderms, such as the origin of the mutable connective tissues. Histologically they are quite similar to collagenous tissues of vertebrates, but having many more binding positions for proteoglucanes (Tipper et al. 2002; Trotter et al. 1994). But these binding positions provided new capacities: the tissues now have the capability to stiffen and relax just by neuronal stimulation and distribution of $Ca^{2+}$ (Hill 2001; Landeira-Fernandez 2001). This means that they can undertake functions of the muscles. Stiffening of the tissues generated a constant pressure on the hydraulic fluid filling of the coelom which could be held for hours. Furthermore the tissues stabilized the shape of the trunk so that the longitudinal muscles could alternately bend the trunk to the one and to the other side; even rotating movements are possible.

Another important point in the echinoderm evolution is the question how the dermal skeleton originated. The recent results that stiffening of MCT is closely related to the distribution of $Ca^{2+}$ (Hill 2001; Landeira-Fernandez 2001) make it likely that under certain pH-conditions the $Ca^{2+}$ could also react with $CO_3^{2-}$ to $CaCO_3$ so that calcareous spiculae could grow in the tissues. This might be supported by the stiffening of the body by the MCT, because if the body wall is not moved for hours small crystals could grow and finally fuse to form spiculae and ossicles. The ossicles are certainly arranged within a capsule like structure. This endoskeleton provides a further stabilization for the entire body shape and allows the body to enlarge. Furthermore, this ossicle-stabilized body construction attains a better potential for fossil preservation, and therefore fossils of echinoderms can be expected only for body constructions that evolved from this ancestral organisation. Although this mechanism has not been elucidated completely, it might be one explanation as to why so many skeletal elements developed.

## Conclusion

Echinoderms are reconstructed here as derived deuterostomes. This results corresponds with the molecular results, and also with the anagenetic implications of the new animal phylogeny (Adoutte et al. 2000). Chordates and tunicates are the sistergroup of the Ambulacraria (entereopneusts, pterobranchs and echinoderms, Metschnikoff 1881) and, within the Ambulacraria, the hemichordates are the sistergroup of the echinoderms.

Of the various echinoderms known from the fossil record and from the Recent, the presented scenario provides a fruitful basis from which to reconstruct evolutionary lineages in more detail. It can be concluded that homalozoans, including the mitrates, cornutes (Stylophora), homoiostelans and homostelans are indeed echinoderms, because they have also have the functional design of an hydraulic skeletal capsule.

## Acknowledgements

I thank the DFG for funding the project GU566/1-1. For many helpful comments I thank Raimund Haude (Göttingen), Manfred Grasshoff (Frankfurt), Stefan Peters (Frankfurt) and Tareq Syed (Frankfurt). I thank Walter Bock (Washington) and Andrew Smith (London) for improving the language and several helpful comment, and Antje Siebel-Stelzner for the illustrations.

Conclusion